\documentclass[10pt]{article}

\usepackage{ol2}
\usepackage[draft]{hyperref}
\usepackage{amsmath}
\usepackage{siunitx}
\newcommand{\um}{\si{\micro\metre}}

\usepackage{caption}
\usepackage{pdfpages}

\renewcommand\author[1]{\noindent%
   \normalsize \bf \begin{center} #1 \end{center}\rm \vskip-1pc}

\renewcommand\address[1]{\noindent%
   \small  \begin{center} \it #1 \end{center}\rm \normalsize \vskip-.3pc}

\begin{document}

\title{Diamond Brillouin Lasers}

\author{Robert J. Williams,$^{1,*}$ Zhenxu Bai,$^{1,2}$ Soumya Sarang,$^{1}$ Ondrej Kitzler,$^{1}$ David J. Spence,$^{1}$ and Richard P. Mildren$^{1}$}

\address{
$^1$MQ Photonics Research Centre, Department of Physics and Astronomy, Macquarie University, NSW 2109, Australia,
\\
$^2$National Key Laboratory of Science and Technology on Tunable Laser, \\Harbin Institute of Technology, Harbin 150001, China \\
$^*$Corresponding author: robert.williams@mq.edu.au
}

\vspace{16pt}

\noindent \textbf{The coherent interaction between optical and acoustic waves via stimulated Brillouin scattering (SBS) is a fundamental tool for manipulating light at GHz frequencies. Its narrowband and noise-suppressing characteristics have recently enabled microwave-photonic functionality in integrated devices based on chalcogenide glasses, silica and silicon \cite{Lee2012,Li2013,Morrison2017,Jiang2016,Casas-Bedoya2015}. Diamond possesses much higher acoustic and bandgap frequencies and superior thermal properties, promising increased frequency, bandwidth and power; however, fabrication of low-loss optical and acoustic guidance structures \cite{VanLaer2015} with the resonances matched to the Brillouin shift \cite{Lee2012} is currently challenging. Here we use intense cavity-enhanced Raman generation to drive a diamond Brillouin laser without acoustic guidance. Our versatile configuration---the first demonstration of a free-space Brillouin laser---provides tens-of-watts of continuous Brillouin laser output on a 71~GHz Stokes shift with user switching between single Stokes and Brillouin frequency comb output. These results open the door to high-power, high-coherence lasers and Brillouin frequency combs, and are a major step towards on-chip diamond SBS devices.}

SBS interactions provide an important bridge between optical and microwave frequencies, enabling generation and signal processing capabilities with high resolution, broad bandwidth and wide tunability that far surpass capabilities of electronic components \cite{Li2013,Jiang2016,Choudhary2017,Liu2017}. The combination of a GHz frequency response and ultra-narrow linewidth (10s of MHz), both of which are widely tunable via the pump spectral properties \cite{Pant2014}, enables microwave processing functionality such as reconfigurable narrow-band filters, phase-shifters and time delays \cite{Choudhary2017,Aryanfar2017,Pagani2014a,Liu2017}. SBS lasers also provide noise suppression via the acoustic field \cite{Debut2000,Behunin2018}, and can be cascaded for ultra-low-noise lasers and frequency combs \cite{Lee2012,Li2013,Kabakova2013,Loh2015,Suh2017,Buettner2014} in microwave synthesis \cite{Li2013} and spectroscopy \cite{Loh2015}. The prospect of integrating these optical synthesis and processing capabilities onto a miniaturized format has stimulated a large effort in on-chip waveguide and resonator-based devices, predominantly in chalcogenide, silica and silicon \cite{Otterstrom2018,Li2013,Choudhary2017,Pant2014,Casas-Bedoya2015}. However, nonlinear absorption, thermal effects and a limited range of Brillouin frequencies (typically 5--20 GHz) in these materials \cite{VanLaer2015,Merklein2016,Wolff2015} limit optical power handling and spectral control---both of which are key to device performance \cite{Li2013}---which has motivated a search for alternative and hybrid material platforms \cite{Morrison2017,Wolff2015,Gundavarapu2017,Wolff2014}.

Diamond's suite of exceptional optical and physical properties make it an outstanding candidate for extreme photonics applications in high-power lasers, quantum optics, bio-photonics and sensing \cite{McKay2017,Bernien2013,Maletinsky2012}. Its high sound velocity and wide bandgap also increase the available Brillouin frequency range to 40--330~GHz \cite{Grimsditch1975}, up to an order of magnitude higher than other SBS materials \cite{Merklein2016,Wolff2014}. Continuous powers and power densities at hundreds of watts and 1~GW.cm$^{-2}$ are routinely sustained without deleterious nonlinear effects \cite{McKay2017,Williams2014,Williams2018} in contrast to silicon, for example, for which values are orders of magnitude lower (10~MW.cm$^{-2}$ at tens of milliwatts) due to the increased role of multi-photon absorption \cite{VanLaer2015}. Thus diamond provides a promising path towards much higher power and power densities, wider wavelength coverage and higher acoustic frequencies, and hence promise for greatly expanding SBS device capability.

SBS is challenging to realize in crystals due to the requirement for either long interaction lengths with optical and acoustic confinement \cite{VanLaer2015,Kittlaus2016,Shin2013} (which also reduces the threshold for parasitic nonlinearities, notably in low-bandgap materials \cite{VanLaer2015,Wolff2015}); or high-Q resonators with precise matching of the cavity mode spacing to the Brillouin frequency shift \cite{Lee2012}. Despite recent progress in the development of photonic devices in diamond \cite{Burek2016,Latawiec2018}, fabrication of low-loss waveguides and resonators remains a challenge due to diamond being extremely hard and inert, and acoustic confinement is made all-the-more difficult in a stiff, high-acoustic-velocity material \cite{VanLaer2015,Poulton2013,Shin2013}.

\begin{figure}[b!]
\centerline{\includegraphics[width=6.2cm]{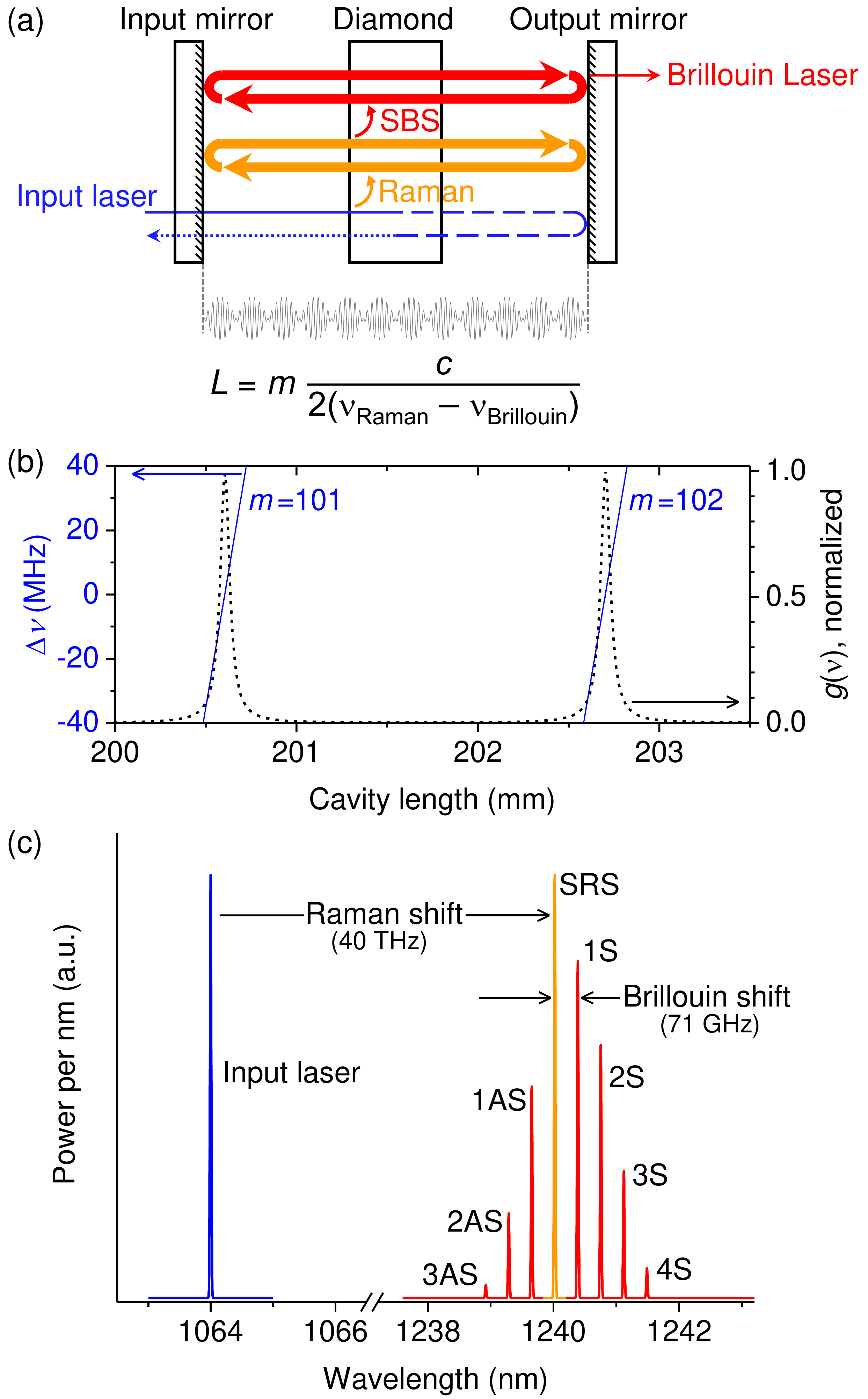}}			
\caption{(a) The Raman-pumped diamond Brillouin laser concept. The Raman field is double-pass pumped by an external 1.06~\um{} multi-mode or single-mode input laser. To simultaneously provide resonance at the lasing Raman wavelength and its corresponding Brillouin shift, the cavity length $L$ is adjusted to satisfy the condition given by the equation below the diagram to within the Brillouin gain linewidth ($m$ is an integer, $c =$ the speed of light, $\nu_{\text{Raman}}$ is the frequency of the lasing Raman mode and $\nu_{\text{Brillouin}}$ is its Brillouin-shifted Stokes frequency). (b) Calculated Brillouin gain $g(\nu)$ as a function of cavity length, assuming a Brillouin gain linewidth of 20~MHz and a pump wavelength of 1240 nm. The blue solid curve shows the detuning $\Delta \nu$ of the closest cavity longitudinal mode from the peak of the Brillouin gain spectrum. (c) Example spectrum of a Brillouin laser frequency comb, including the input laser spectrum (blue curve), Raman pump and Brillouin laser output (orange and red curves, respectively), highlighting the first-Stokes Raman shift from the 1064~nm input laser accompanied by several Brillouin Stokes (1S, 2S, etc.) and anti-Stokes shifts (1AS, etc.).}
\label{fig:fig1}
\end{figure}

We have overcome these barriers by using an intense Raman-generated pump field in a free-space diamond laser. In this approach, stimulated Raman scattering (SRS) efficiently transfers power from an external input laser beam to a pump field resonant with a high-finesse cavity (Fig.~\ref{fig:fig1}). Pump intensities exceeding 100~MW.cm$^{-2}$ are generated with mode sizes more than a hundred times the acoustic wavelength, enabling Brillouin lasing without major diffraction loss of the acoustic wave from the optical mode. The cavity length is adjusted to provide control over the output spectrum according to the resonance conditions for the Raman pump and Brillouin laser fields. The high thermal conductivity and damage threshold of diamond enables a much greater range of parameter space to be explored compared to other solid-state candidates for SBS.

\begin{figure}[htb]
\centerline{\includegraphics[width=12cm]{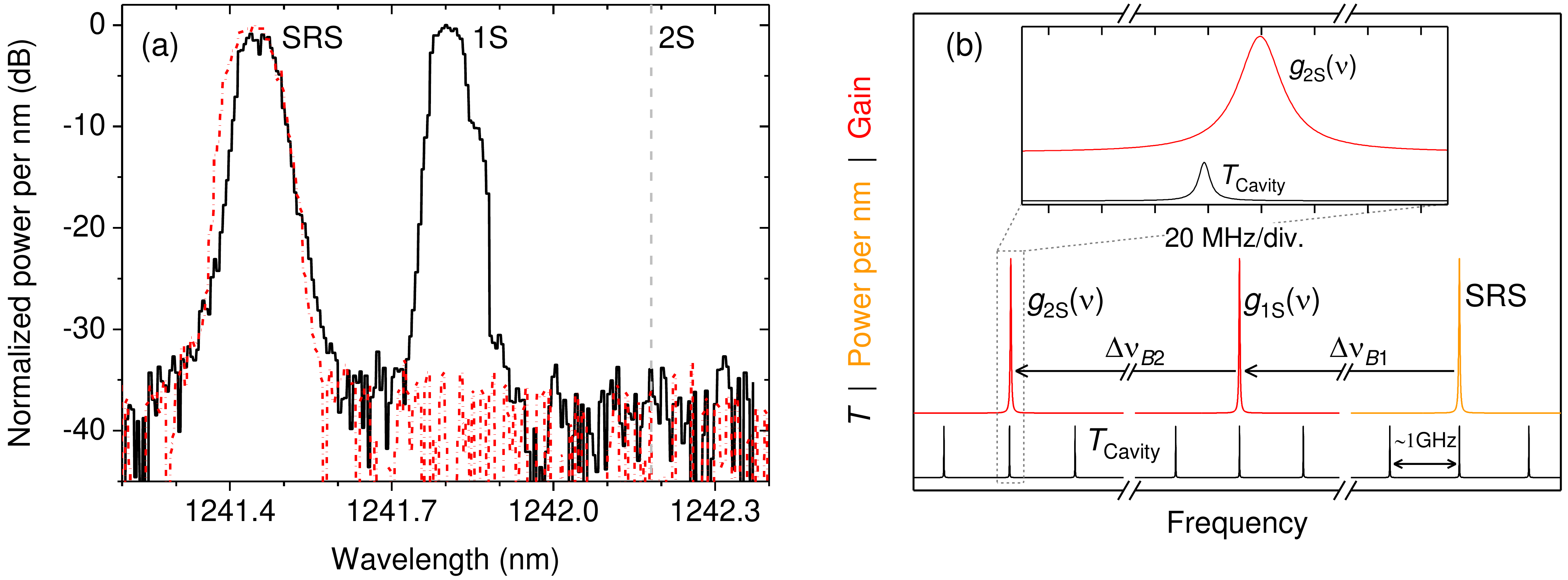}}			
\caption{(a) Output spectrum of the SLM-pumped diamond Brillouin laser with cavity length tuned onto resonance (black solid curve) and off resonance (red dashed curve). The difference in cavity length between the two cases shown is less than 200~\um{}. The dashed vertical line indicates the wavelength corresponding to a second Stokes Brillouin shift, which was not observed. (b) shows the frequency separation between the second-Stokes Brillouin gain line and the cavity modes (illustrated here with a 20~MHz Brillouin gain linewidth and 5~MHz cavity mode linewidth).}
\label{fig:SLM}
\end{figure}

To coherently pump the acoustic wave, we first exploited the spatial-hole-burning-free nature of Raman gain to generate a single longitudinal mode (SLM) Raman field as the Brillouin laser pump in a standing-wave cavity \cite{Lux2016}. By setting the cavity length to a double-resonance at the Raman and Brillouin-shifted wavelengths (Fig.~\ref{fig:fig1}a,b), Brillouin lasing was observed as a distinct peak at 1241.79~nm in addition to the Raman laser line at 1241.42~nm (Fig.~\ref{fig:SLM}). The frequency separation of 71 $\pm$ 3~GHz corresponds to the back-scattered Brillouin frequency shift from the longitudinal acoustic mode in diamond propagating along $\langle 110 \rangle$ \cite{Williams2015,Grimsditch1975}. Scanning the mirror spacing over several hundred microns tunes a cavity resonance over the Brillouin gain profile (Fig.~\ref{fig:fig1}b), providing selectable enhancement or complete suppression of Brillouin lasing. Further experimental details are provided in the Methods. Single-Stokes output was observed without cascading to higher orders, even for cases where the power in the Brillouin mode exceeded that of the Raman pump. Since the Brillouin frequency shift is proportional to the pump frequency ($\Delta \nu_B = 2 n v_a \nu_0 / c$) \cite{Chiao1964} the frequency spacing decreases with Stokes order, an effect exacerbated in diamond due to its high product of sound velocity $v_a$ and index $n$. The calculated walk-off of the second-Stokes frequency from the cavity resonance is 21~MHz, thus it can be deduced from the absence of second-Stokes that the Brillouin linewidth of diamond is not more than a few times this value. A gain linewidth of a few tens of MHz is in line with values measured for other materials (TeO$_2$, 27~MHz \cite{Dubinskii2004}; fused silica, 41~MHz \cite{Faris1993}; silicon, 30--40~MHz \cite{VanLaer2015}; and chalcogenide glasses, 13--34~MHz \cite{Pant2011}). 

\begin{figure}[t!]
\centerline{\includegraphics[width=5.8cm]{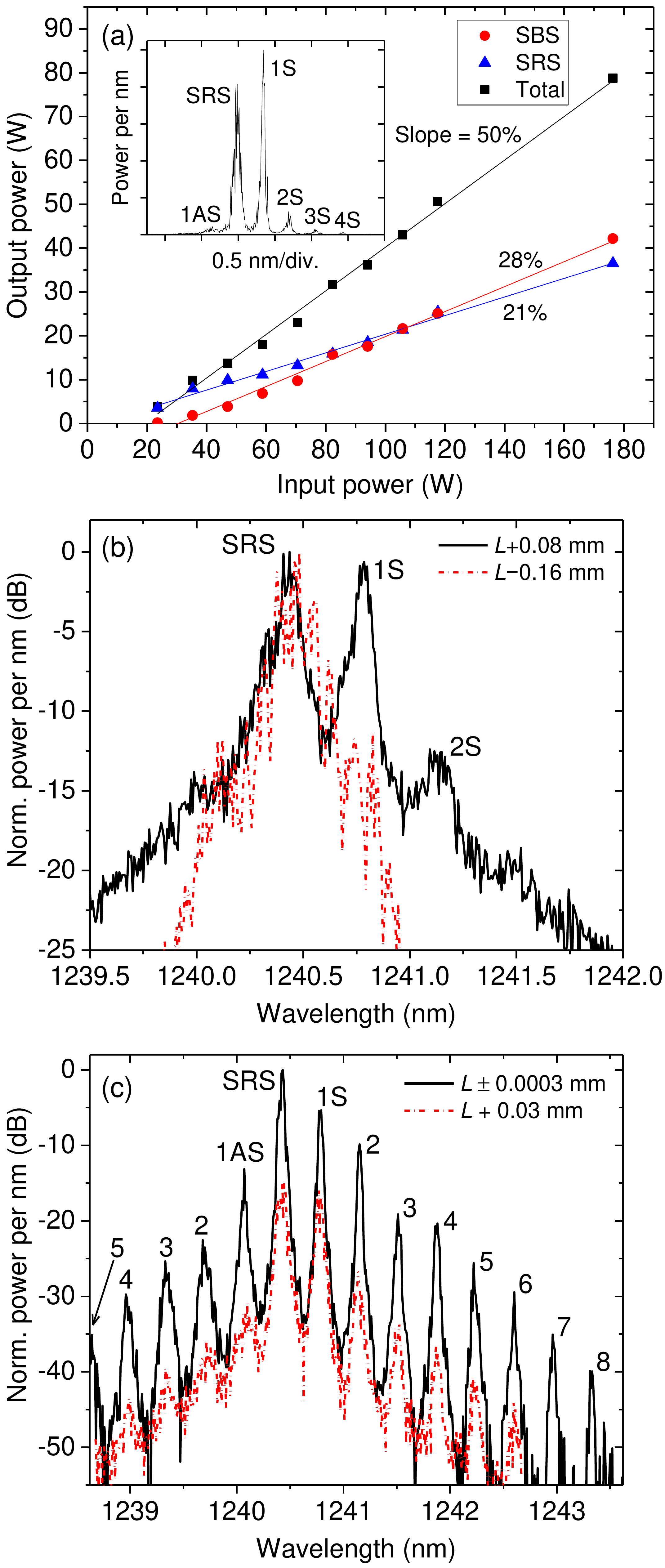}}
\caption{(a) Power characterisation of the MLM-pumped diamond Brillouin laser. The combined Raman and Brillouin output (black squares) had a slope efficiency of 50\% and reached 79~W for 176~W input beam power. Red circles show the total output at all Brillouin Stokes and anti-Stokes frequencies, and blue triangles indicate the power output in the Raman modes. The power in the first Stokes Brillouin component was also quantified, and averaged 38\% of the total output power for input beam powers above 50~W. The inset shows the laser output spectrum at 176~W input power on a linear scale. (b) and (c) show the output spectrum for off- and on-resonance for comb generation, i.e. as a function of detuning of the cavity longitudinal-mode spectrum from the Brillouin gain peak (here $L$ is the optimum cavity length for comb generation). (b) compares one case optimized for conversion to the first Brillouin Stokes wavelength (black solid curve) with another case at 0.16~mm cavity adjustment away from Brillouin resonance, showing Raman conversion but suppressed SBS. (c) shows the spectrum obtained with adjustment of the cavity length near to resonance for cascaded Brillouin Stokes and anti-Stokes generation (red dashed curve), and the spectrum obtained after further sub-micron cavity-length adjustment optimized for comb generation (black solid curve). Note both curves are normalized and on a logarithmic scale, and the red dashed curve is offset by $-15$~dB for clarity. The more finely-tuned Brillouin frequency comb exhibits distinct comb lines up to the 8th Stokes order and the 5th anti-Stokes order, with approximately 10~dB greater suppression of power-per-nm between comb lines. In each case the observed spectrum was stable over several minutes, and repeatable by adjustment of the cavity length.}
\label{fig:MLM}
\end{figure}

We also investigated Brillouin lasing pumped by a broadened Raman field comprising of approximately 21 longitudinal modes (spanning $\sim 20$~GHz). This was achieved by using a multi-longitudinal-mode (MLM) 1064~nm input beam for driving the Raman field. The Brillouin laser spectrum in this case consisted of multiple Stokes orders with power up to 42.2~W (Fig.~\ref{fig:MLM}a) across all Stokes orders and 31.0 W for the first Stokes order. End-mirror translation again provided control over the Brillouin laser spectrum, including operating regimes consisting of single and cascaded SBS lines and completely suppressed Brillouin lasing (Fig.~\ref{fig:MLM}b,c). For the cavity length providing the widest comb, up to 8 Brillouin Stokes shifts and 5 anti-Stokes shifts were observed (Fig.~\ref{fig:MLM}c). The presence of anti-Stokes lines is indicative of four-wave-mixing (FWM), which potentially provides a mechanism for phase-locking and mode-locked pulse generation \cite{Buettner2016}. The comb width was found to be highly sensitive to mirror spacing, with sub-wavelength tuning leading to notable sharpening of each spectral component with reduced spectral power density between each comb line (Fig.~\ref{fig:MLM}c). We expect that such behaviour is analogous to that seen in other Brillouin \cite{Buettner2016} and Kerr frequency comb systems \cite{Li2012a} in which cavity or pump tuning leads to the onset of phase-locking with subsequent comb broadening and noise reduction \cite{Li2012a,Saha2013}. Thus, these results highlight potential for generating high-power, high-repetition rate mode-locked pulses and low-noise frequency combs from diamond Brillouin lasers with active cavity stabilization.

The Brillouin comb in Fig.~\ref{fig:MLM}c contains anti-Stokes lines and a triangular profile consistent with Kerr FWM \cite{Dong2016}. Kerr FWM generates equally-spaced comb lines that are resonant with the cavity, and is phase-matched for co-propagating fields in the standing-wave configuration \cite{Buettner2016}. A Kerr-dominated generation mechanism is also consistent with the diminished role of cascaded Brillouin gain as a result of the frequency walk-off from cavity resonances (Fig.~\ref{fig:SLM}b). Kerr comb generation is expected to be enhanced in the presence of multiple pairs of Raman and Brillouin-shifted modes (each of which is resonant with the cavity) due to higher peak intensities from longitudinal mode beating. This explains the observation of comb generation containing a first-Stokes Brillouin component with only a fraction of the intensity of the Raman pump (see Fig.~\ref{fig:MLM}c), in contrast to the SLM-pumped case. However, it is anticipated that SLM combs generated by Kerr FWM will be observed at higher input powers.

While our Raman-pumped Brillouin laser approach provides advantages for attaining laser threshold in a bulk format, there are good prospects for further reducing threshold and achieving lasing in other arrangements. Currently the observed threshold leads to a calculated gain coefficient that is much lower than expected from reported photo-elastic coefficients and the acoustic linewidth deduced above. The threshold for the SLM-pumped case (800~W of intracavity power), for example, yields a Brillouin gain coefficient of 0.7~cm/GW (see Supplementary Material). A similar value is obtained for the multimode case, even under the conservative assumption that all the Raman power is contained in a single mode. However, a calculation based on the photoelastic tensor for diamond \cite{Grimsditch1975} ($p^2 = 0.303$ for our configuration, see Supplementary Material) and for a linewidth in the range 10--50~MHz yields a Brillouin gain coefficient between 50 and 300~cm/GW. This range is at the high end of all other materials \cite{Bai2018,Faris1993}, comparable only to chalcogenide glasses and TeO$_2$ in solid state. We attribute the high threshold in the present experiments to the interactive nature of the Raman and Brillouin gain in the shared cavity. Broadening in the Raman longitudinal mode spectrum is expected under the influence of depletion from a lasing Brillouin field and becomes unstable for the present case where the Raman field has a shorter response time compared to the Brillouin ($T_2^{\text{Raman}} \approx T_2^{\text{Brillouin}}/1000$; see Supplementary Material). These complex dynamics may be avoided by employing externally pumped architectures with potential for greatly reduced thresholds.

The ability for diamond to be continuously pumped at high power with resilience to stress fracture and thermal lensing \cite{Mildren2013a}, even in the presence of impurity absorption, provides an excellent medium for exploring a wide range of Brillouin phenomena. Our laser results highlight great potential for high power lasers and amplifiers with intrinsically low quantum defect and narrow linewidth. Our demonstrated control of the output spectrum foreshadows the attractive prospect of selectable frequency generation to single Stokes or multiple Stokes and widely-spaced frequency combs. Power-scalable single-Stokes output is a critical factor for low-noise Brillouin lasers \cite{Behunin2018} which in our case is aided by the large frequency walk-off per cascade. There are also prospects for transfering concepts from pulsed SBS in liquids and gases---such as phase-conjugate mirrors \cite{Carr1985} and beam combination \cite{Bai2018,Wang2007}---into all-solid-state designs and into the continuous-wave regime. Finally, this work comprises a major step towards diamond on-chip Brillouin devices, in which ultrahigh Brillouin frequencies, high Brillouin gain and the robust material platform are anticipated to combine to greatly extend the capabilities of integrated Brillouin photonics.

\section*{Methods}

\subsection*{SLM-pumped diamond Brillouin laser}

The input laser was a 50~W ytterbium-doped fiber amplifier (IPG Photonics, YAR-LPSF) seeded by an external-cavity diode laser (TOPTICA Photonics, DL-DFB-BFY), operating at 1065~nm with a linewidth of 5~MHz. The input beam was focussed to a waist radius of 36~\um{} in the diamond using a 50~mm focal-length lens. The diamond Brillouin laser configuration consisted of a near-concentric linear resonator with the diamond centered on the waist of the cavity mode. The diamond was a $1.2 \times 4 \times 8$~mm$^3$ sample with low birefringence and low nitrogen content ($<20$~ppb, ElementSix Ltd.), with beam propagation along $\langle 110 \rangle$ and parallel to the 8 mm-long edge. Anti-reflection coatings optimized for 1240~nm were applied to the diamond end facets. These coatings have shown ability to withstand extremely high continuous-wave power densities ($>4$~MW/cm linear power density \cite{Williams2014,Williams2018}) due to their direct contact with such a thermally conductive substrate. The plane-concave input coupler mirror had a 50~mm radius of curvature and was optimized for high-reflectivity at 1240~nm and high transmission at 1064~nm; while the plane-concave output coupler mirror had a 75~mm radius of curvature and was optimized for high-reflectivity at 1240~nm and 1064~nm, in order to double-pass the input beam through the diamond. Consequently, the cavity losses at the Stokes wavelengths (Raman and Brillouin) were dominated by diamond losses of about 0.5\% per round trip. The total cavity length was approximately 126~mm, and the Stokes mode radius in the diamond was 65~\um{}. In this configuration, we restricted the injected input power to less than 25~W to avoid damage to the output coupler mirror, which did not have a very high laser-induced damage threshold at 1240~nm. The mount for the output coupler mirror was fixed to a piezo-controlled translation stage, enabling sub-micron adjustment of the cavity length.

The spectral content of the diamond lasers (both SLM and MLM-pumped) were characterized using an optical spectrum analyser (Anritsu, MS9710A) with spectral resolution of 0.076~nm at 1240~nm.

\subsection*{MLM-pumped laser}

The input laser was a quasi-CW-pumped Nd:YAG laser, operating at 1064 nm with 0.2~ms pulses and a repetition frequency of 40~Hz. The motivation for using this mode of operation is to simplify development of input laser sources with multi-hundred-watt power levels. As detailed and demonstrated in our previous diamond Raman laser papers \cite{Williams2015,Williams2014}, this pulse duration is orders of magnitude longer than the time-scale for diamond to reach a thermal equilibrium equivalent to continuous-wave operation with the diamond side faces cooled to room temperature. Since thermal effects occur on far longer time-scales than any other relevant optical or acousto-optical effects, the laser behaviour is effectively CW. On-time (i.e. CW) power levels of both input and Stokes light were obtained from oscilloscope traces of the laser output captured with fast photodiodes, which were calibrated in power from average power measurements obtained with thermal power meters. The diamond laser cavity setup was almost identical to that of the SLM-pumped diamond Brillouin laser, except that the IC and OC mirrors had radii of curvature of 100~mm and 50~mm, respectively, the OC mirror had 0.5\% transmission at 1240~nm, and the input beam focussing lens had focal length 100~mm. The total cavity length, Stokes mode radius in the diamond and input beam waist radius in the diamond were 152~mm, 62~\um{} and 45~\um{}, respectively.

\section*{Acknowledgements}

This work was supported by the Australian Research Council
(DP150102054), and the U.S. Air Force Research Laboratory (FA2386-15-1-4075).



\section*{Supplementary material}

# Supplementary Material

**Calculation of the Brillouin gain coefficient from the photoelastic coefficient**

The Brillouin gain coefficient $g_0$ is dependent on the photoelastic coefficient $p$ according to [1, 2]

$$g_0 = \frac{n^7 p^2 \omega_s^2}{\rho v_a c^3 \Gamma}, \qquad (1)$$

where $n$ is the refractive index, $\omega_s$ is the Stokes angular frequency, $\rho$ is the mass density, $v_a$ is the acoustic velocity (18.3 km/s for the longitudinal mode propagating along $\langle 110 \rangle$ [3]), $c$ is the vacuum speed of light and $\Gamma$ is the Brillouin gain linewidth (in radians/s). In crystals, $p$ is dependent on the polarization of the pump and Stokes fields and the propagation direction of the acoustic phonons, and is found using

$$p^2 = (e_s T e_p)^2, \qquad (2)$$

where $e_s$ and $e_p$ are Stokes and pump polarization vectors, respectively, and $T$ is the Brillouin scattering tensor. For the case of longitudinal acoustic phonons propagating along $\langle 110 \rangle$ in diamond, $T$ is given by [3]:

$$T = \begin{bmatrix} p_{11} + p_{12} & 2p_{44} & 0 \\ 2p_{44} & p_{11} + p_{12} & 0 \\ 0 & 0 & 2p_{12} \end{bmatrix}, \qquad (3)$$

where $p_{ij}$ are the photoelastic constants of diamond. A map of $p^2$ values as a function of linear pump and Stokes polarizations is shown in Fig. S1. The maximum Brillouin gain occurs for pump and Stokes polarizations aligned perpendicular to [001] (i.e. max. gain is for horizontal polarization in our setup), with a $p^2$ value of 0.303. Using this value for $p^2$ and an estimated range for the Brillouin gain linewidth $\Gamma/2\pi = 20$–60 MHz, we calculate a likely range for the Brillouin gain coefficient $g_0 = 52$–311 cm/GW.

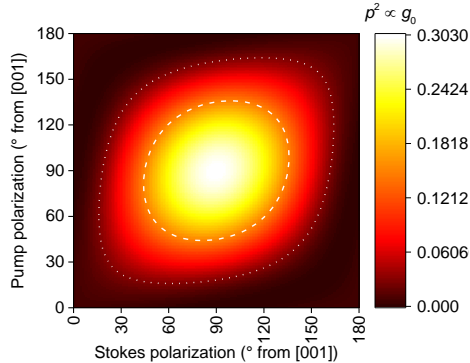

Fig. S1: Map of $p^2$ as a function of pump and Brillouin Stokes linear polarization angles (measured from the vertical, [001]) for Brillouin backscattering via the longitudinal acoustic mode propagating parallel to $\langle 110 \rangle$. Note that here "pump" refers to the field that pumps the Brillouin interaction, which in the case of our experiments is a Raman Stokes field. The white dashed (inner) and dotted (outer) contour lines indicate where $p^2$ is equal to half the maximum value and 10% of the maximum value, respectively. Polarization effects in diamond Raman lasers have been elucidated in other work [4].

**Calculation based on the SLM-pumped experimental results**

An estimation of the Brillouin gain coefficient was determined experimentally for the narrow-linewidth-pumped laser under the assumption that the Raman pump comprised of a stable SLM. The condition

for Brillouin laser threshold was derived by adapting an existing free-space Raman laser model [5]. After accounting for the fact that the Stokes field only experiences Brillouin gain from a counter-propagating pump (whereas Raman gain is equal for co- and counter-propagating pump and Stokes beams), as well as the fact that the circulating Raman field makes a double pass through the crystal per round trip, the threshold condition for Brillouin lasing can be expressed as

$$P_{\text{Th}_B} = \frac{(T + 2\alpha L)\Lambda}{4g_B \arctan(\xi)}, \tag{4}$$

where $P_{\text{Th}_B}$ is the circulating pump power in the cavity at threshold for Brillouin lasing (i.e. the circulating intra-cavity power of the Raman field), $T$ is the transmission of the output coupler mirror (approximately 0.2% accounting for coating losses), $\alpha = 0.003$ cm$^{-1}$ is the loss per centimetre in the diamond at 1240 nm, $L = 0.8$ cm is the diamond length, $g_B$ is the Brillouin gain coefficient, and $\Lambda$ and $\xi$ are parameters that account for the effect of focussed beams on the overall gain. Expressions for $\Lambda$ and $\xi$ are given in Eq. 8 and 9 of [5], but since in our case the Raman and Brillouin fields both occupy fundamental transverse modes of the same resonator and are at effectively equal wavelengths, they simplify to

$$\Lambda_B = \frac{\lambda_S}{n}, \quad \text{and} \tag{5}$$

$$\xi_B = \frac{L}{2z_S}, \tag{6}$$

where $\lambda_S = 1240$ nm, $z_S$ is the Rayleigh length of the Raman and Brillouin modes, and $n = 2.42$ is the refractive index of diamond.

In order to calculate the circulating Raman intra-cavity power at Brillouin threshold, we use the aforementioned Raman laser model and the measured threshold for Raman lasing (19.1 W of injected 1064 nm input beam power). The input beam and Raman Stokes waist radii in the diamond were 36 µm and 65 µm, respectively, and both beams had $M^2 = 1$. Since for the Raman case, the input and Stokes wavelengths and beam parameters are significantly different, the simplifications above ($\Lambda = \Lambda_B$ and $\xi = \xi_B$) do not apply. Here it is convenient to express $\Lambda$ and $\xi$ in terms of the waist radii and Rayleigh lengths as

$$\Lambda_R = \frac{\pi}{2}\sqrt{(w_i^2 + w_S^2)\left(\frac{w_i^2}{z_i^2} + \frac{w_S^2}{z_S^2}\right)}, \quad \text{and} \tag{7}$$

$$\xi_R = \frac{L}{2}\sqrt{\left(\frac{w_i^2}{z_i^2} + \frac{w_S^2}{z_S^2}\right)/(w_i^2 + w_S^2)}, \tag{8}$$

where $w_i$, $w_S$, $z_i$ and $z_S$ are the waist radii and Rayleigh lengths of the input (1064 nm) and Raman Stokes beams, respectively. In the case of the Raman laser, for which the 1064 nm input beam also double-passes through the crystal, the condition for Raman lasing threshold is then

$$P_{\text{Th}_R} = \frac{(T + 2\alpha L)\Lambda_R}{4(1 + R_i)g_R \arctan(\xi_R)}, \tag{9}$$

where $g_R = 10$ cm/GW and $R_i$ is the reflectivity of the output-coupler mirror at 1064 nm (approximately 0.97 to account for diamond coating losses at 1064 nm on a double pass). Using the parameters given above, this gives $P_{\text{Th}_R} = 19.1$ W in agreement with the experimentally measured value. Subsequently, using the same model [5] we calculate the intra-cavity circulating power of the Raman field at the 1064 nm input power of 22.4 W, corresponding to the observed threshold for Brillouin lasing. This model gives an expression for the input power required to produce a circulating intra-cavity Raman Stokes power as

$$P_p = \frac{T + 2\alpha L}{\eta} P_S^{\text{circ}} \left(1 - \exp\left[-2(1 + R_i)GP_S^{\text{circ}}\right]\right)^{-1}, \tag{10}$$

where $G = 2g_R \arctan(\xi)/(\eta\Lambda)$ is the Raman power gain in the focussed geometry, $P_S^{\text{circ}}$ is the circulating intra-cavity power of the Raman field, and $\eta = 1064/1240$ is the quantum defect of the Raman shift. For an input power of 22.4 W we calculate $P_S^{\text{circ}} = 800$ W. Substituting this value for $P_{\text{Th}_B}$ into Eq. (4), we calculate $g_B = 0.7$ cm/GW. Uncertainties in the diamond loss coefficient, diamond coating losses and mirror coating losses give rise to significant uncertainty in this value (a factor of two to four would be reasonable) since in this case the output coupler is nominally HR; however, the uncertainty in this value is insignificant compared to the discrepancy between it and the estimated value from the photoelastic tensor and Brillouin linewidth (as given in the previous section), for the reasons discussed in the paper and in the following section.

**Effect of broadening and instability in the Raman longitudinal-mode spectrum on the observed threshold**

The much-higher observed Brillouin lasing threshold than expected from estimations of the gain coefficient can be attributed to spectral broadening and instability in the Raman laser field that pumps the Brillouin interaction in diamond. This is similar to spectral broadening effects that have been observed and analysed in intra-cavity-pumped Raman lasers [6, 7], and briefly outlined below.

In an intra-cavity-pumped Raman laser, a Raman crystal shares an optical cavity with a fundamental inversion-based laser crystal and the fundamental laser field pumps the Raman laser field via SRS. In such lasers, the Raman laser field provides an intra-cavity loss for the fundamental field, with a finite spectral width determined by the Raman gain linewidth. Since the Raman gain linewidth is typically narrower than the gain profile of the fundamental laser field, the fundamental laser field tends to broaden with increasing pump power above the Raman laser threshold, which in turn reduces the efficiency of conversion from the fundamental field to the Raman laser field. This has been mitigated in practise by using intra-cavity filtering elements (etalons) to eliminate spectral broadening of the fundamental laser field [6].

The same mechanism applies in our Brillouin laser, causing broadening (or instability, as outlined below) in the intra-cavity Raman field which pumps the Brillouin laser. However, in our case a single Brillouin longitudinal mode only provides a loss for a single longitudinal mode of the Raman field because the Brillouin linewidth is far narrower than the longitudinal mode spacing of the cavity. Neighbouring longitudinal modes for the Raman field have almost identical gain to the initial Raman mode, and can easily reach threshold, since the Raman gain bandwidth is 45 GHz in diamond—more than 40 times the longitudinal mode spacing. Therefore, at the very onset of Brillouin lasing, the corresponding Raman longitudinal mode is clamped (due to conversion via SBS, much like in a two-Stokes Raman laser [8]), and neighbouring longitudinal modes in the Raman line will reach threshold and deplete energy from the input laser field, suppressing further growth of the Brillouin mode. Once these neighbouring modes grow to a sufficient intensity for Brillouin lasing, they too will drive more Brillouin modes and repeat the process, giving rise to additional Raman modes without significant additional growth in the Brillouin output power. In this way, instead of increasing Brillouin output power, the onset of Brillouin lasing gives rise to additional Raman lasing modes and broadening of the Raman output spectrum. Since the differential in gain between neighbouring longitudinal Raman modes is so small, we would expect negligible growth in the Brillouin laser field until the Raman line is broadened to about the Raman gain linewidth (45 GHz), in the case of a steady-state system.

While the above modelling assumes all fields are in a steady state, it is likely that our laser is operating in a dynamic regime where the same effect on Brillouin output power can be realized without severe broadening of the Raman spectrum. Since the Raman phonon lifetime (7 ps [9]) is approximately three orders of magnitude shorter than the Brillouin phonon lifetime, the Raman field can respond much faster (e.g. in build-up time) than the Brillouin field. Since the Brillouin field provides such a narrow-band loss on the Raman field as to only affect a single longitudinal mode, it is likely that the Raman lasing field would respond by rapidly mode-




\begin{thebibliography}{10}
\newcommand{\enquote}[1]{``#1''}

\bibitem{Lee2012}
H.~Lee, T.~Chen, J.~Li, K.~Y. Yang, S.~Jeon, O.~Painter, and K.~J. Vahala,
  \enquote{Chemically etched ultrahigh-{Q} wedge-resonator on a silicon chip,}
  \natp \textbf{6}, 369--373 (2012).

\bibitem{Li2013}
J.~Li, H.~Lee, and K.~J. Vahala, \enquote{Microwave synthesizer using an
  on-chip {B}rillouin oscillator,} \natc \textbf{4}, 3097 (2013).

\bibitem{Morrison2017}
B.~Morrison, A.~Casas-Bedoya, G.~Ren, K.~Vu, Y.~Liu, A.~Zarifi, T.~G. Nguyen,
  D.-Y. Choi, D.~Marpaung, S.~J. Madden, A.~Mitchell, and B.~J. Eggleton,
  \enquote{Compact {B}rillouin devices through hybrid integration on silicon,}
  Optica \textbf{4}, 847--854 (2017).

\bibitem{Jiang2016}
H.~Jiang, D.~Marpaung, M.~Pagani, K.~Vu, D.-Y. Choi, S.~J. Madden, L.~Yan, and
  B.~J. Eggleton, \enquote{Wide-range, high-precision multiple microwave
  frequency measurement using a chip-based photonic {B}rillouin filter,} Optica
  \textbf{3}, 30--34 (2016).

\bibitem{Casas-Bedoya2015}
A.~Casas-Bedoya, B.~Morrison, M.~Pagani, D.~Marpaung, and B.~J. Eggleton,
  \enquote{Tunable narrowband microwave photonic filter created by stimulated
  {B}rillouin scattering from a silicon nanowire,} \ol \textbf{40}, 4154--4157
  (2015).

\bibitem{VanLaer2015}
R.~Van~Laer, B.~Kuyken, D.~Van~Thourhout, and R.~Baets, \enquote{Interaction
  between light and highly confined hypersound in a silicon photonic nanowire,}
  \natp \textbf{9}, 199--203 (2015).

\bibitem{Choudhary2017}
A.~Choudhary, B.~Morrison, I.~Aryanfar, S.~Shahnia, M.~Pagani, Y.~Liu, K.~Vu,
  S.~Madden, D.~Marpaung, and B.~J. Eggleton, \enquote{Advanced integrated
  microwave signal processing with giant on-chip {B}rillouin gain,} \jlt
  \textbf{35}, 846--854 (2017).

\bibitem{Liu2017}
Y.~Liu, A.~Choudhary, D.~Marpaung, and B.~J. Eggleton, \enquote{Gigahertz
  optical tuning of an on-chip radio frequency photonic delay line,} Optica
  \textbf{4}, 418--423 (2017).

\bibitem{Pant2014}
R.~Pant, D.~Marpaung, I.~V. Kabakova, B.~Morrison, C.~G. Poulton, and B.~J.
  Eggleton, \enquote{On-chip stimulated {B}rillouin scattering for microwave
  signal processing and generation,} \lpr \textbf{8}, 653--666 (2014).

\bibitem{Aryanfar2017}
I.~Aryanfar, D.~Marpaung, A.~Choudhary, Y.~Liu, K.~Vu, D.-Y. Choi, P.~Ma,
  S.~Madden, and B.~J. Eggleton, \enquote{Chip-based {B}rillouin radio
  frequency photonic phase shifter and wideband time delay,} \ol \textbf{42},
  1313--1316 (2017).

\bibitem{Pagani2014a}
M.~Pagani, D.~Marpaung, and B.~J. Eggleton, \enquote{Ultra-wideband microwave
  photonic phase shifter with configurable amplitude response,} \ol
  \textbf{39}, 5854--5857 (2014).

\bibitem{Debut2000}
A.~Debut, S.~Randoux, and J.~Zemmouri, \enquote{Linewidth narrowing in
  {B}rillouin lasers: Theoretical analysis,} \pra \textbf{62}, 023803 (2000).

\bibitem{Behunin2018}
R.~Behunin, N.~T. Otterstrom, P.~T. Rakich, S.~Gundavarapu, and D.~J.
  Blumenthal, \enquote{Fundamental noise dynamics in cascaded-order {B}rillouin
  lasers,} Preprint at http://arXiv.org/abs/1802.03894  (2018).

\bibitem{Kabakova2013}
I.~V. Kabakova, R.~Pant, D.-Y. Choi, S.~Debbarma, B.~Luther-Davies, S.~J.
  Madden, and B.~J. Eggleton, \enquote{Narrow linewidth {B}rillouin laser based
  on chalcogenide photonic chip,} \ol \textbf{38}, 3208--3211 (2013).

\bibitem{Loh2015}
W.~Loh, A.~A.~S. Green, F.~N. Baynes, D.~C. Cole, F.~J. Quinlan, H.~Lee, K.~J.
  Vahala, S.~B. Papp, and S.~A. Diddams, \enquote{Dual-microcavity
  narrow-linewidth {B}rillouin laser,} Optica \textbf{2}, 225--232 (2015).

\bibitem{Suh2017}
M.-G. Suh, Q.-F. Yang, and K.~J. Vahala, \enquote{Phonon-limited-linewidth of
  {B}rillouin lasers at cryogenic temperatures,} \prl \textbf{119}, 143901
  (2017).

\bibitem{Buettner2014}
T.~F.~S. B\"uttner, M.~Merklein, I.~V. Kabakova, D.~D. Hudson, D.-Y. Choi,
  B.~Luther-Davies, S.~J. Madden, and B.~J. Eggleton, \enquote{Phase-locked,
  chip-based, cascaded stimulated {B}rillouin scattering,} Optica \textbf{1},
  311--314 (2014).

\bibitem{Otterstrom2018}
N.~T. Otterstrom, R.~O. Behunin, E.~A. Kittlaus, Z.~Wang, and P.~T. Rakich,
  \enquote{A silicon {B}rillouin laser,} Science \textbf{360}, 1113--1116
  (2018).

\bibitem{Merklein2016}
M.~Merklein, A.~Casas-Bedoya, D.~Marpaung, T.~F.~S. B\"uttner, M.~Pagani,
  B.~Morrison, I.~V. Kabakova, and B.~J. Eggleton, \enquote{Stimulated
  {B}rillouin scattering in photonic integrated circuits: Novel applications
  and devices,} \jstqe \textbf{22}, 336--346 (2016).

\bibitem{Wolff2015}
C.~Wolff, P.~Gutsche, M.~J. Steel, B.~J. Eggleton, and C.~G. Poulton,
  \enquote{Power limits and a figure of merit for stimulated {B}rillouin
  scattering in the presence of third and fifth order loss,} \opex \textbf{23},
  26628--26638 (2015).

\bibitem{Gundavarapu2017}
S.~Gundavarapu, M.~Puckett, T.~Huffman, R.~Behunin, J.~Wu, T.~Qiu, G.~M.
  Brodnik, C.~Pinho, D.~Bose, and P.~T. Rakich, \enquote{Integrated waveguide
  {B}rillouin laser,} Preprint at http://arXiv.org/abs/1709.04512  (2017).

\bibitem{Wolff2014}
C.~Wolff, R.~Soref, C.~G. Poulton, and B.~J. Eggleton, \enquote{Germanium as a
  material for stimulated {B}rillouin scattering in the mid-infrared,} \opex
  \textbf{22}, 30735--30747 (2014).

\bibitem{McKay2017}
A.~McKay, D.~J. Spence, D.~W. Coutts, and R.~P. Mildren, \enquote{Diamond-based
  concept for combining beams at very high average powers,} \lpr \textbf{11},
  1600130 (2017).

\bibitem{Bernien2013}
H.~Bernien, B.~Hensen, W.~Pfaff, G.~Koolstra, M.~S. Blok, L.~Robledo, T.~H.
  Taminiau, M.~Markham, D.~J. Twitchen, L.~Childress, and R.~Hanson,
  \enquote{Heralded entanglement between solid-state qubits separated by three
  metres,} Nature \textbf{497}, 86--90 (2013).

\bibitem{Maletinsky2012}
P.~Maletinsky, S.~Hong, M.~S. Grinolds, B.~Hausmann, M.~D. Lukin, R.~L.
  Walsworth, M.~Lon\v{c}ar, and A.~Yacoby, \enquote{A robust scanning diamond
  sensor for nanoscale imaging with single nitrogen-vacancy centres,} \natn
  \textbf{7}, 320--324 (2012).

\bibitem{Grimsditch1975}
M.~H. Grimsditch and A.~K. Ramdas, \enquote{Brillouin scattering in diamond,}
  \prb \textbf{11}, 3139--3148 (1975).

\bibitem{Williams2014}
R.~J. Williams, O.~Kitzler, A.~McKay, and R.~P. Mildren, \enquote{Investigating
  diamond {R}aman lasers at the 100 {W} level using quasi-continuous-wave
  pumping,} \ol \textbf{39}, 4152--4155 (2014).

\bibitem{Williams2018}
R.~J. Williams, O.~Kitzler, Z.~Bai, S.~Sarang, H.~Jasbeer, A.~McKay,
  S.~Antipov, A.~Sabella, O.~Lux, D.~J. Spence, and R.~P. Mildren,
  \enquote{High power diamond {R}aman lasers,} \jstqe \textbf{24}, 1602214
  (2018).

\bibitem{Kittlaus2016}
E.~A. Kittlaus, H.~Shin, and P.~T. Rakich, \enquote{Large {B}rillouin
  amplification in silicon,} \natp \textbf{10}, 463--467 (2016).

\bibitem{Shin2013}
H.~Shin, W.~Qiu, R.~Jarecki, J.~A. Cox, R.~H. Olsson, A.~Starbuck, Z.~Wang, and
  P.~T. Rakich, \enquote{Tailorable stimulated {B}rillouin scattering in
  nanoscale silicon waveguides,} \natc \textbf{4}, 1944 (2013).

\bibitem{Burek2016}
M.~J. Burek, J.~D. Cohen, S.~M. Meenehan, N.~El-Sawah, C.~Chia, T.~Ruelle,
  S.~Meesala, J.~Rochman, H.~A. Atikian, M.~Markham, D.~J. Twitchen, M.~D.
  Lukin, O.~Painter, and M.~Lon\v{c}ar, \enquote{Diamond optomechanical
  crystals,} Optica \textbf{3}, 1404--1411 (2016).

\bibitem{Latawiec2018}
P.~Latawiec, V.~Venkataraman, A.~Shams-Ansari, M.~Markham, and M.~Lon\v{c}ar,
  \enquote{Integrated diamond {R}aman laser pumped in the near-visible,} \ol
  \textbf{43}, 318--321 (2018).

\bibitem{Poulton2013}
C.~G. Poulton, R.~Pant, and B.~J. Eggleton, \enquote{Acoustic confinement and
  stimulated {B}rillouin scattering in integrated optical waveguides,} \josab
  \textbf{30}, 2657--2664 (2013).

\bibitem{Lux2016}
O.~Lux, S.~Sarang, O.~Kitzler, D.~J. Spence, and R.~P. Mildren,
  \enquote{Intrinsically stable high-power single longitudinal mode laser using
  spatial hole burning free gain,} Optica \textbf{3}, 876--881 (2016).

\bibitem{Williams2015}
R.~J. Williams, J.~Nold, M.~Strecker, O.~Kitzler, A.~McKay, T.~Schreiber, and
  R.~P. Mildren, \enquote{Efficient {R}aman frequency conversion of high-power
  fiber lasers in diamond,} \lpr \textbf{9}, 405--411 (2015).

\bibitem{Chiao1964}
R.~Y. Chiao, C.~H. Townes, and B.~P. Stoicheff, \enquote{Stimulated {B}rillouin
  scattering and coherent generation of intense hypersonic waves,} \prl
  \textbf{12}, 592--595 (1964).

\bibitem{Dubinskii2004}
M.~A. Dubinskii and L.~D. Merkle, \enquote{Ultrahigh-gain bulk solid-state
  stimulated {B}rillouin scattering phase-conjugation material,} \ol
  \textbf{29}, 992--994 (2004).

\bibitem{Faris1993}
G.~W. Faris, L.~E. Jusinski, and A.~P. Hickman, \enquote{High-resolution
  stimulated {B}rillouin gain spectroscopy in glasses and crystals,} \josab
  \textbf{10}, 587--599 (1993).

\bibitem{Pant2011}
R.~Pant, C.~G. Poulton, D.-Y. Choi, H.~McFarlane, S.~Hile, E.~Li, L.~Thevenaz,
  B.~Luther-Davies, S.~J. Madden, and B.~J. Eggleton, \enquote{On-chip
  stimulated {B}rillouin scattering,} \opex \textbf{19}, 8285--8290 (2011).

\bibitem{Buettner2016}
T.~F.~S. B\"uttner, C.~G. Poulton, M.~J. Steel, D.~D. Hudson, and B.~J.
  Eggleton, \enquote{Phase-locking in cascaded stimulated {B}rillouin
  scattering,} \njp \textbf{18}, 025003 (2016).

\bibitem{Li2012a}
J.~Li, H.~Lee, T.~Chen, and K.~J. Vahala, \enquote{Low-pump-power,
  low-phase-noise, and microwave to millimeter-wave repetition rate operation
  in microcombs,} \prl \textbf{109}, 233901 (2012).

\bibitem{Saha2013}
K.~Saha, Y.~Okawachi, B.~Shim, J.~S. Levy, R.~Salem, A.~R. Johnson, M.~A.
  Foster, M.~R.~E. Lamont, M.~Lipson, and A.~L. Gaeta, \enquote{Modelocking and
  femtosecond pulse generation in chip-based frequency combs,} \opex
  \textbf{21}, 1335--1343 (2013).

\bibitem{Dong2016}
M.~Dong and H.~G. Winful, \enquote{Unified approach to cascaded stimulated
  {B}rillouin scattering and frequency-comb generation,} \pra \textbf{93},
  043851 (2016).

\bibitem{Bai2018}
Z.~Bai, H.~Yuan, Z.~Liu, P.~Xu, Q.~Gao, R.~J. Williams, O.~Kitzler, R.~P.
  Mildren, Y.~Wang, and Z.~Lu, \enquote{Stimulated {B}rillouin scattering
  materials, experimental design and applications: A review,} \om \textbf{75},
  626--645 (2018).

\bibitem{Mildren2013a}
R.~P. Mildren, A.~Sabella, O.~Kitzler, D.~J. Spence, and A.~M. McKay,
  \emph{Diamond {R}aman Laser Design and Performance} (Wiley-VCH Verlag,
  Weinheim, Germany, 2013), pp. 239--276.

\bibitem{Carr1985}
I.~D. Carr and D.~C. Hanna, \enquote{Performance of a {Nd:YAG}
  oscillator/ampflifier with phase-conjugation via stimulated brillouin
  scattering,} \apb \textbf{36}, 83--92 (1985).

\bibitem{Wang2007}
S.~Wang, Z.~LÜ, D.~Lin, L.~E.~I. Ding, and D.~Jiang, \enquote{Investigation of
  serial coherent laser beam combination based on {B}rillouin amplification,}
  \lpb \textbf{25}, 79--83 (2007).

\end{thebibliography}
\end{document}